# Bit Parallel 6T SRAM In-memory Computing with Reconfigurable Bit-Precision

Kyeongho Lee[1], Jinho Jeong[1], Sungsoo Cheon[1], Woong Choi[2], and Jongsun Park[1]
[1] School of Electrical Engineering, Korea University, Seoul, Korea
[2] School of Electronics Engineering, Sookmyung Women's University, Seoul, Korea
{rudgh0143, k007312, chssis, jongsun}@korea.ac.kr, woongchoi@sookmyung.ac.kr

## ABSTRACT

This paper presents 6T SRAM cell-based bit-parallel in-memory computing (IMC) architecture to support various computations with reconfigurable bit-precision. In the proposed technique, bit-line computation is performed with a short WL followed by BL boosting circuits, which can reduce BL computing delays. By performing carry-propagation between each near-memory circuit, bit-parallel complex computations are also enabled by iterating operations with low latency. In addition, reconfigurable bit-precision is also supported based on carry-propagation size. Our 128KB in/near memory computing architecture has been implemented using a 28nm CMOS process, and it can achieve 2.25GHz clock frequency at 1.0V with 5.2% of area overhead. The proposed architecture also achieves 0.68, 8.09 TOPS/W for the parallel addition and multiplication, respectively. In addition, the proposed work also supports a wide range of supply voltage, from 0.6V to 1.1V.

## 1 INTRODUCTION

In processor-centric architecture, the separation between computing units and memories generates numerous data movement through data bus, causing huge energy consumption. For the last decade, an abrupt increase of data-intensive applications has made the data movement issue become one of the largest bottlenecks in computing system. As processor-centric architecture is not very efficient to perform data-centric applications, such as deep learning and real-time visual/streaming processing, reducing data movement between computing units and memories, has been an active research topic.

As one of the research efforts to relieve this data movement burden, processing in memory (PIM) or in memory computing (IMC) has been proposed [1]-[15]. The main idea of PIM or IMC is to perform some or all of the computations within or near memories, so that it can reduce the data movement between computing units and memories. In [1], row-wise operation has been proposed using the reconfiguration of 6T SRAM cell. This work can perform simple logic operation, such as bit-wise logic operation. However, this work is inappropriate for high-performance data-centric applications since it has to repeat the simple bit-wise operations to perform complex computation. In [2], 8T transposable bit-cell based bit-serial IMC architecture has been proposed to support more complex operations such as shift, add, and multiplication. This work nonetheless had to endure high latency, slow clock frequency, and large cell area overhead. Moreover, the techniques employed to prevent read disturbance issue inevitably limits the memory performance. This contradicts the ultimate goal of IMC which is to reduce the data movement while satisfying the system performance. In [5], the approach to improve the performance of in-memory addition while avoiding read disturbance issue has been covered. Although it employed 6T SRAM based IMC for area efficiency, it still had a drawback in terms of data access. Furthermore, when implementing the application which requires computation in diverse bit-precision, such as machine learning inference, the limited bit-precision architecture can result in unnecessary use of hardware. Therefore, the need for flexibility of bit-precision must also be considered.

In this paper, to address the complexity of operations, performance of memory, and flexible bit-precision, we present a 6T SRAM based bit-parallel in-memory computation architecture. In the proposed approach, the bit parallel computations, which are specialized in carry propagations in adding/multiplying operation, are executed inside memories with low latency. In order to enhance the memory clock frequency while avoiding read disturbance during BL computing, a short WL pulse with a BL boosting scheme is also proposed. Furthermore, to improve the performance, transmission gate based adders are adopted, providing simple bit-wise operations. For the same reason, write-back delay is reduced by separating dummy arrays to be used independently. In addition, by adjusting the size of carry propagation, the additions and multiplications with reconfigurable bit-width are also supported in the proposed architecture. As a result, the proposed 6T SRAM cell based memory banks can provide considerably faster processing speed with small cell area and small peripherals.

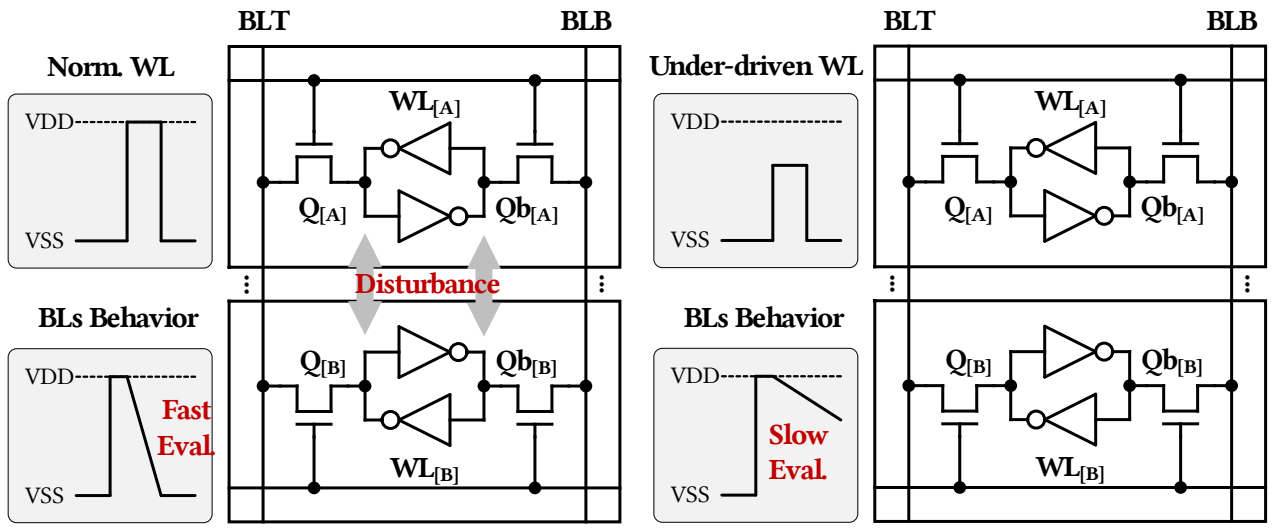

**Figure 1: Conventional BL computing schemes.**

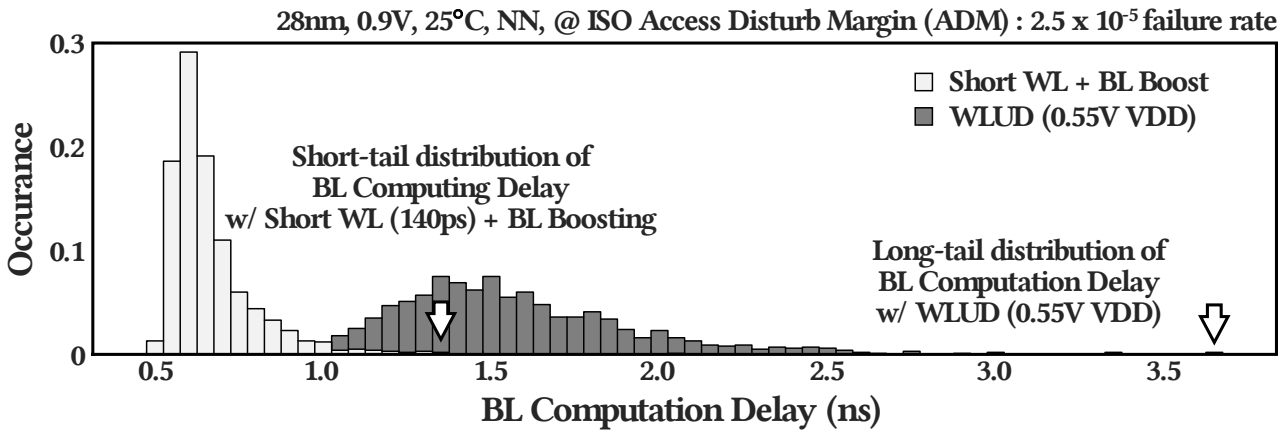

**Figure 2: Comparison BL computation delay between WLUD and Short-WL + BL Boosting**

The rest of this paper is organized as follows. In Section 2, the background of IMC and previous IMC works are introduced. Section 3 describes the proposed bit-parallel in/near-memory computing architecture for low-latency arithmetic operation with reconfigurable bit-precision. In Section 4, the experimental results, using the 28nm CMOS process based post-layout simulation, are presented with the comparisons to previous IMC solutions. Finally, Section 5 concludes this work.

# 2 BACKGROUND

## 2.1 Bit-line Computing

Many previous in-memory computing solutions are based on bit-line computation while activating two WL simultaneously. However, there is a read disturbance issue in the SRAM cells when WLs are activated, as shown in Fig. 1. For example, in the case of 'A=0', 'B=1', BL/BLB are discharged through each cell's ground node. Due to the discharged BL from '1' to '0', the stored data in the cell which held '1' may be flipped. To solve this problem, one of the SRAM assist techniques, Word-Line Under-Drive (WLUD) is employed. However, the weakened access transistor can make BL discharge slow. This eventually affects the memory operation frequency thus leading to performance degradation. Fig. 2 shows the distribution of BL computation delay of the proposed IMC compared to the conventional, 0.55V WL driven 6T SRAM IMC, which has the same read failure rate of $2.5 \times 10^{-5}$. A short pulse based fully driven WL shows smaller deviation. Therefore, in this work, instead of adopting WLUD, we propose fully driven WL, but for a short amount of time, to guarantee the performance and avoid the read disturbance at the same time.

## 2.2 Related Works

To alleviate the data movement of modern computing system, many IMC solutions have been proposed so far. A simple bit-wise logic operation is implemented using reconfiguration of 6T SRAM cells in [1]. However, this work can only implement functions restricted in their complexity. To resolve this limitation, more recent solutions [2]-[5] offer more complex bit-operations, such as shift, add, and multiplication. An architecture capable of bit-serial arithmetic operation using 8T transposable bit cell is presented in [2], where vector-based operations can be performed. Due to the bit-serial architecture, high-complexity operation can be completed, and high throughput is achieved. However, the approach suffers from large latency as multiplication takes $N^2$ cycles. Additionally, the solution employed for the read disturbance problem incurs slow BL discharge because of the low voltage WL. By adopting ripple-carry-addition using pipelining with latches in array peripheral circuits, faster operation cycles without read disturbance issue could be achieved in [4]. Since this work employed 10T SRAM cells with 2 separate read ports to resolve the problem, low cell efficiency was inevitably followed. Improvement of cell efficiency without read disturbance issue was conducted in [5], with bit-parallel in-memory operations using separate read BL with 6T SRAM cell, where Manchester addition can be performed inside memory with small carry propagation delay. However, it still requires additional area overhead for each local array group, therefore the cell array efficiency remains problematic. Considering data-centric applications such as deep learning processors, it is also important to support different operation bit-widths.

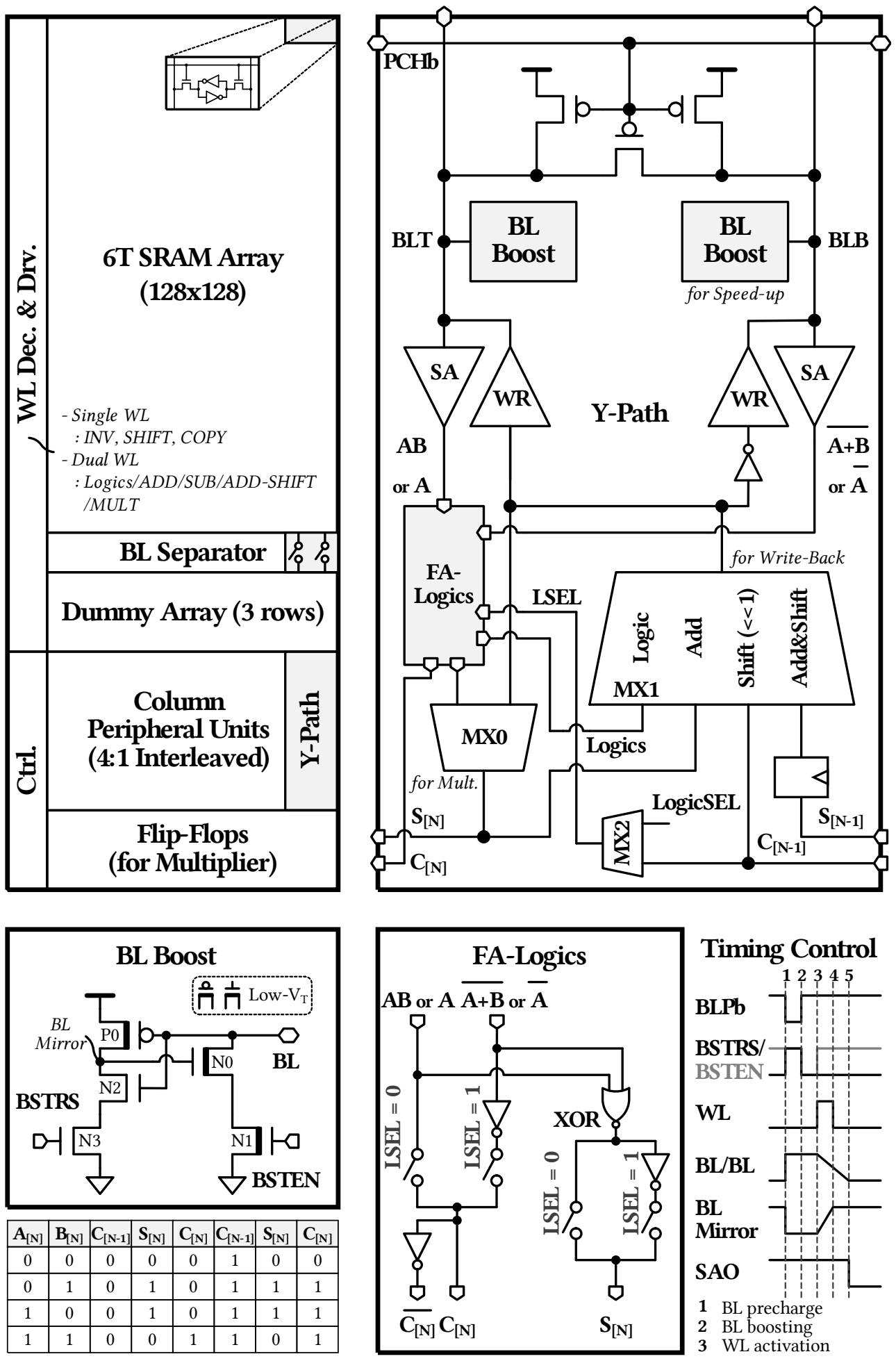

| A[N] | B[N] | C[N-1] | S[N] | C[N] | C[N-1] | S[N] | C[N] |
|---|---|---|---|---|---|---|---|
| 0 | 0 | 0 | 0 | 0 | 1 | 0 | 0 |
| 0 | 1 | 0 | 1 | 0 | 1 | 1 | 1 |
| 1 | 0 | 0 | 1 | 0 | 1 | 1 | 1 |
| 1 | 1 | 0 | 0 | 1 | 1 | 0 | 1 |

**Figure 3: Th proposed 6T SRAM-based bit parallel in-memory-computing architecture.**

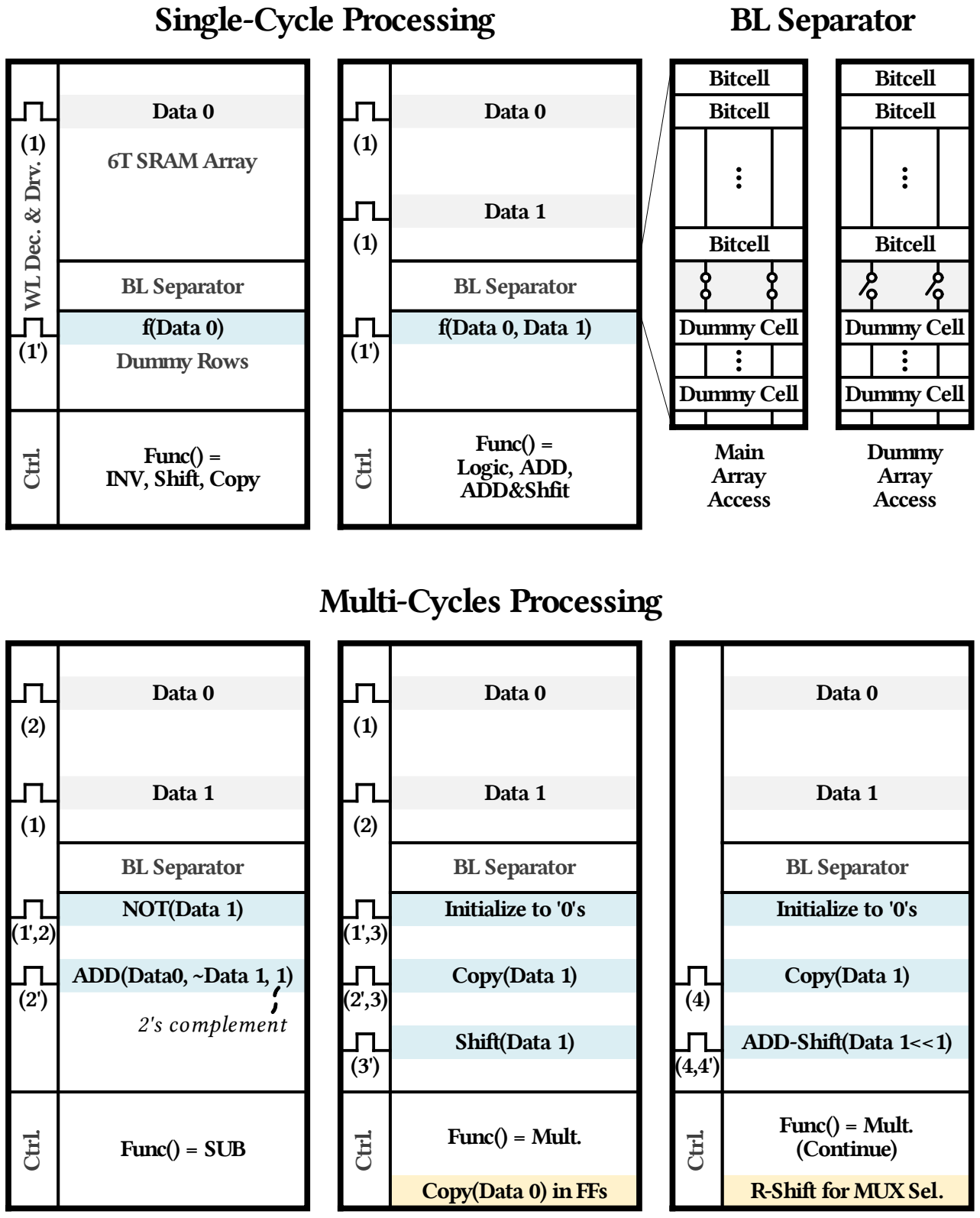

**Figure 4: Operation principles of the proposed IMC.**

Overall in the previous works [1]-[7], each has different drawback in either cell area efficiency, computation complexity, or memory clock frequency. In the following, we present 6T SRAM based bit-parallel in-memory computing architecture, in which read disturbance and performance degradation issues are addressed together using a short WL followed by BL boosting. In addition, the proposed bit-parallel in-memory computing approach can process complex operation with low latency. Details of the proposed IMC architecture will be presented in section III.

# 3 PROPOSED IN-MEMORY COMPUTING

## 3.1 Operation Principles

Fig. 3 illustrates the architecture of the proposed in-memory-processing cells and peripherals to provide low latency arithmetic operation with reconfigurable bit-precision. As shown in this figure, the proposed architecture includes additional computing blocks, that can be summarized to i) bit-line (BL) separator and dummy array in the SRAM cell array area, and ii) BL booster, transmission gate based full-adder and logic gates (FA-Logics)

**Table 1: Th  supported operations and cycles**

| Type | Operation | Cycles | Type | Operation | Cycles |
|---|---|---|---|---|---|
| Logic | NAND/AND | 1 | Integer | ADD | 1 |
| | NOR/OR | 1 | | SUB | 2 |
| | XNOR/XOR | 1 | | MULT | N+2 |
| | NOT, Shift (<<1) | 1 | | ADD-Shift | 1 |

* N represents the data bit-width

with three multiplexers and bunch of flip-flops in the column peripheral area. To avoid area overhead caused by modified cell or memory array, the conventional 6T SRAM cell and the array structure are employed. In the proposed architecture, BL boosting scheme and transmission gate based FA-Logics reduce the computation delay with small additional area overhead (5.2% of array area). Moreover, to perform complex computation which requires iterative operation, such as multiplication, we use dummy array and few flip-flops.

The proposed in-memory-processing can be divided into i) BL computation, and ii) logic and arithmetic operations. For the BL computation, the detailed procedure is as follows: Firstly, while the BL precharges, BL boosting scheme is initialized by boosting the reset (BSTRS) signal, which makes BL mirror node set to VSS. Next, when the WLs are activated, if the computation result is 'Low', BL is discharged slightly. Accordingly, P0 transistor is gradually turned on, which leads BL mirror node to go high. Thus, BL will be discharged through N0-N1, which has larger discharge path than that of SRAM cell. As shown in BL Boost of Fig. 3, to catch-up the small BL swing, caused by short WL pulse-width to prevent the read disturbance, low threshold voltage (LVT) device is used for P0, N0, and N1 transistors. Otherwise, if the BL computation result is 'High', BL boosting scheme is not activated. After the acceleration of the BL swing, the single-ended sense amplifiers (SAs) generate the BL computation results. At this time, depending on the number of activated WL, the SA outputs are set to AB/~(A+B) or A/~A. Here, the A and B indicate the accessed memory data in the SRAM array.

For the simple logic operations, which is not included in the BL computation results, an OR gate and three inverters with four transmission gates (illustrated as switches) are embedded in the 'FA-Logics'. As shown in 'Y-Path' and 'FA-Logics' of Fig. 3, by using the selection signal of multiplexer MX2 and 'LogicSEL' signal, all the logic operations can be generated. For the simple arithmetic operations, such as addition (ADD) and subtraction (SUB), the logic results can be mixed cost-effectively in the proposed in-memory-computing feature. The Boolean equations of the FA outputs, e.g. sum and carry-out, can be expressed as:

$$S_{[N]} = C_{[N-1]}(A_{[N]} \odot B_{[N]}) + \sim C_{[N-1]}(A_{[N]} \oplus B_{[N]}) \quad (1)$$

$$C_{[N]} = C_{[N-1]}(A_{[N]} | B_{[N]}) + \sim C_{[N-1]}(A_{[N]} \& B_{[N]}) \quad (2)$$

As shown in (1)-(2) and 'FA-Logics' of Fig. 3, by using the carry-out of the right-side 'Y-Path' ($C_{[N-1]}$) as switch control signal, the BL computation results (AB and ~(A+B)) can be converted to the FA outputs. For the shift and add-and-shift operations, this carry-propagation path is used by controlling the multiplexers. In case of 1-bit shift operation with single-WL activation, the 'FA-Logics' outputs the original data (A) to the C[N] node, and the MX1 selects the write-back data as the propagated data (C[N-1]). On the other hand, in case of the add-shift operation, the sum value of 'FA-Logics' is passed to the MX0 and propagated to the left-side 'Y-Path'. During write-back, the flip-flop releases the propagated sum value (S[N-1]) to perform the shift operation. Since this propagation path is not only used for ADD and SUB operations, but also used for shift and add-and-shift operations, the overhead of the complex arithmetic operation, such as multiplication (MULT), is alleviated in the proposed in-memory-computing. The detailed procedure of the MULT operation and comparison in various bit-precision cases will be discussed in Section 3.2 and Section 4, respectively.

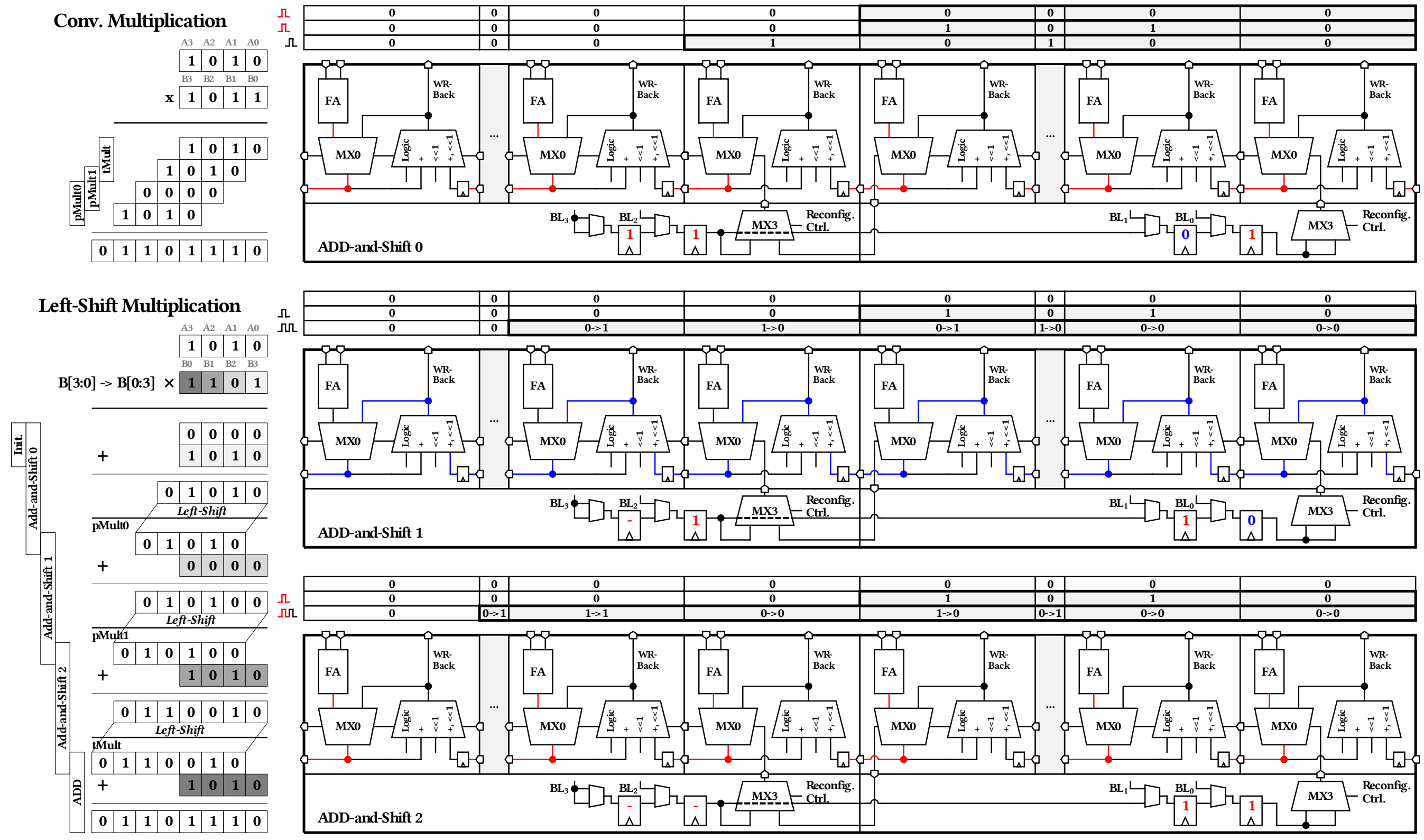

**Figure 5: Multiplication procedure of the proposed in-memory-computing architecture.**

For the SUB operation, the overall procedure is presented in the bottom-left of Fig. 4. As shown in this figure, the SUB operation is started with a NOT operation. At this time, the SRAM read path outputs the stored data (Data 1 in Fig. 4), and the column peripheral units write-back the inverted 'Data 1' to the dummy array. Please note that, in the Fig. 4, the numbers below the WL pulse indicate the accessing cycle, and the prime symbol (') separates the read and write-back operations. After the NOT operation (including write-back), an ADD operation is performed to subtract the 'Data 1' from the 'Data 0' using two's complement format. As shown in Fig. 4, in order to reduce the write-back power consumption, the BL separator adaptively disconnects the large capacitive BL in the SRAM array. The supported in-memory-computing operations are categorized in Fig. 4 and Table I. As shown in Table I, except for the SUB, and MULT operation, all the other operations only take 1 cycle.

## 3.2 Bit-Parallel Multiplication with Reconfigurable Bit-precision

As shown in the top-left illustration of Fig. 5, the conventional multiplication consists of i) partial products generation, and ii) summation of all the partial products. When these operations are performed as is, 6 (1+2+3) shift operations and 3 addition operations are required for 4×4 multiplication. In the proposed in-memory-computing architecture, by using the 'add-and-shift' function and adopting the left-shift based multiplication algorithm, the multiplication operation can be performed cost-effectively. The left-shift based multiplication is presented in the bottom-left of Fig. 5. As shown in this figure, the reversed multiplicand (B[3:0] → B[0:3]) makes the accumulation of partial products to the left-shift based addition operation. As shown in the '1010×1011' example of Fig. 5, the partial products accumulations in the left-shift based multiplication perform reversely compared to the conventional approach (pMult0 → pMult1 → tMult). At this time, the reversed accumulations can be divided into each add-and-shift operation groups by inserting an initial '0000' to the adding operation. In the proposed in-memory-computing architec-ture, since the add-and-shift function is performed in a single-cycle while minimizing data movement, the multiplication overhead is significantly alleviated compared to the conventional multiple shift and add combination. The detailed hardware operation principle is also illustrated in Fig. 5. During initialization, '0's are written to the first row of dummy array, while multiplier data (B[3:0] = 1011) is stored in the flip-flops. After that, multiplicand data (A[3:0] = 1010) is copied to the second row of dummy array. Here, these steps are omitted in Fig. 5. Next, first and second rows are activated for the add-and-shift operation. During those add-and-shift operations, the propagated data from lower bit to upper bit is selected depending on the flip-flop output. As shown in Fig. 5, in case of the add-and-shift 0 phase (top-right illustration of Fig. 5), since the flip-flop output is '1', the FA output is selected and wrote back, shifted. On the other hand, in case of add-and-shift 1 phase, since the valid data are stored in second and third rows of dummy array, the corresponding WLs are activated. At this time, since the flip-flop output is '0', the data which was propagated from lower bit and stored in the flip-flop of 'Y-Path' is delivered to the write-back path. Similarly, during the add-and-shift 2 phase, the second and third rows are activated to add the previously accumulated data (010100 in Fig. 5) to the multiplicand (1010 in Fig. 5). For the final accumulation of the partial products, the ADD operation is performed to complete the multiplication result. For the NxN multiplication, the total number of cycles can be computed with 2 initialization steps and N iterative add-and-shift operation.

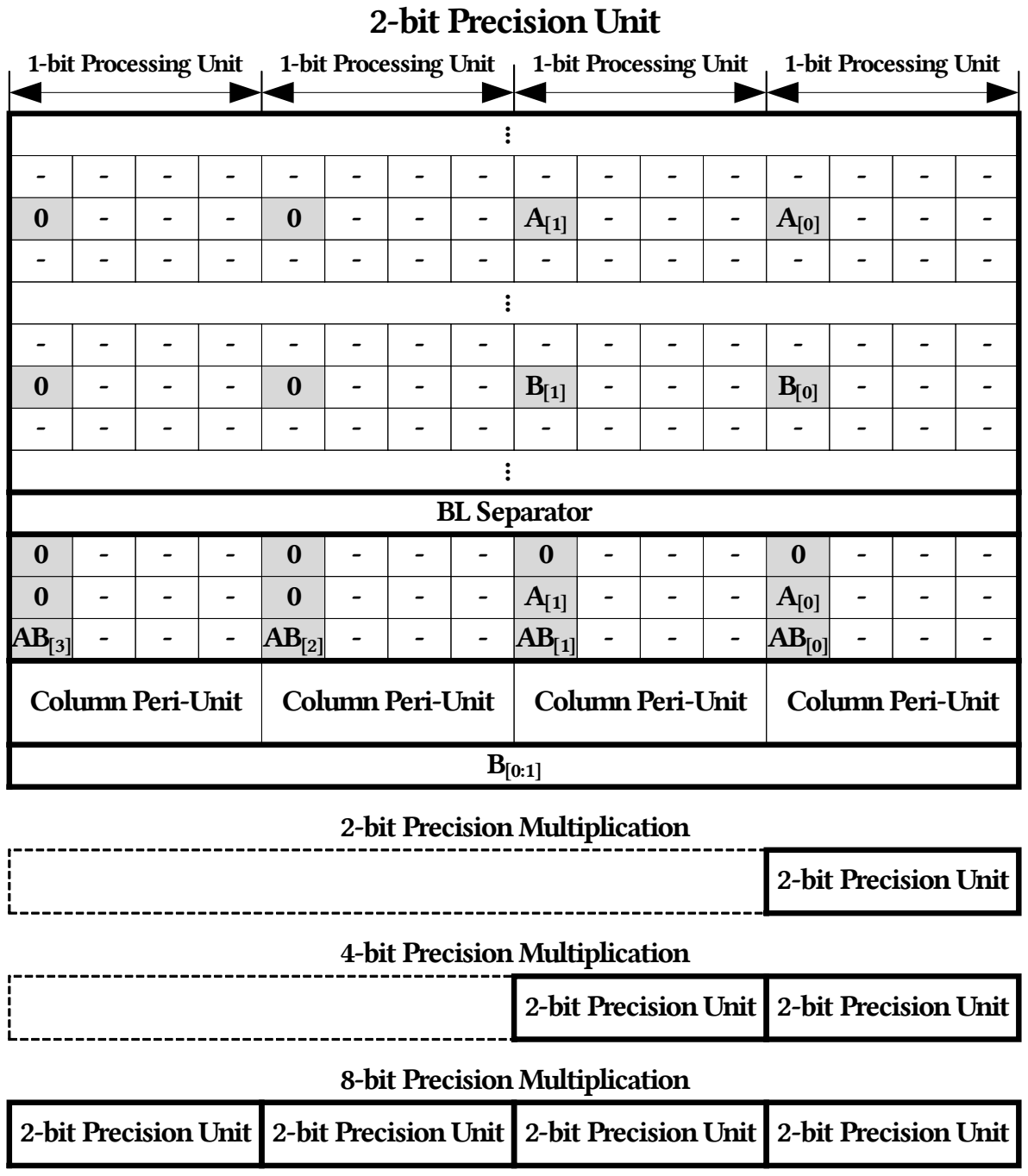

**Figure 6: Reconfigurable bit precision for multiplication.**

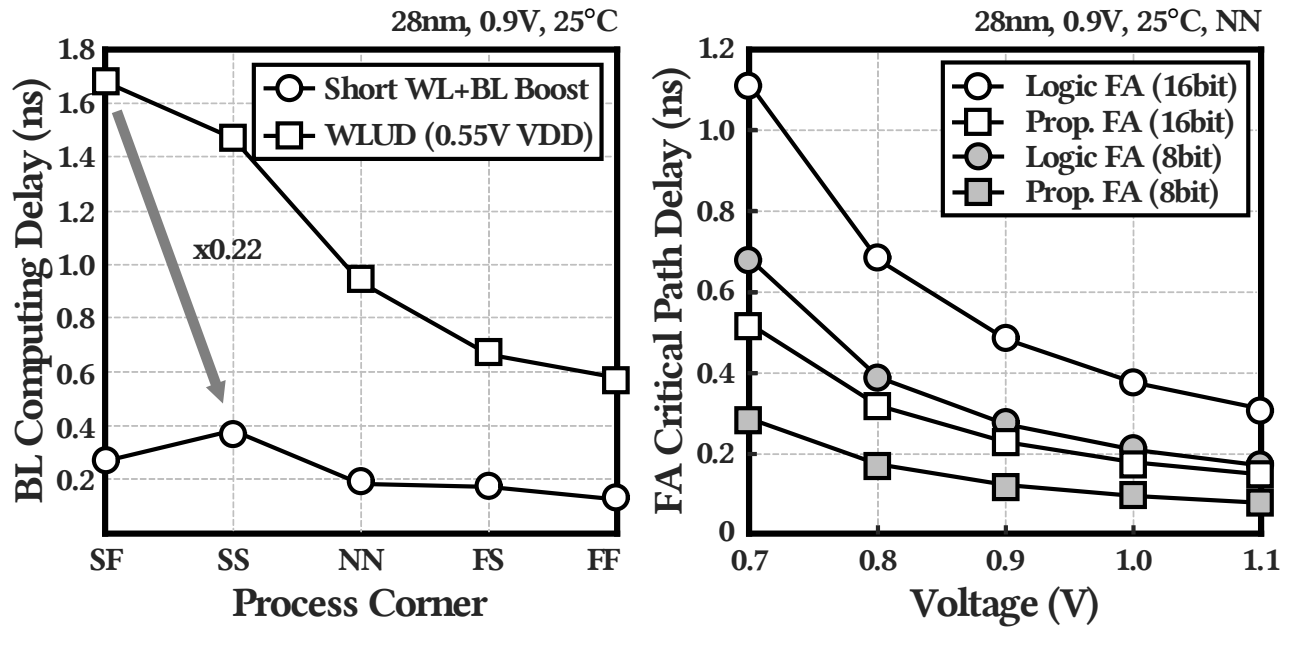

(a) (b)

**Figure 7: Delay Comparisons.**

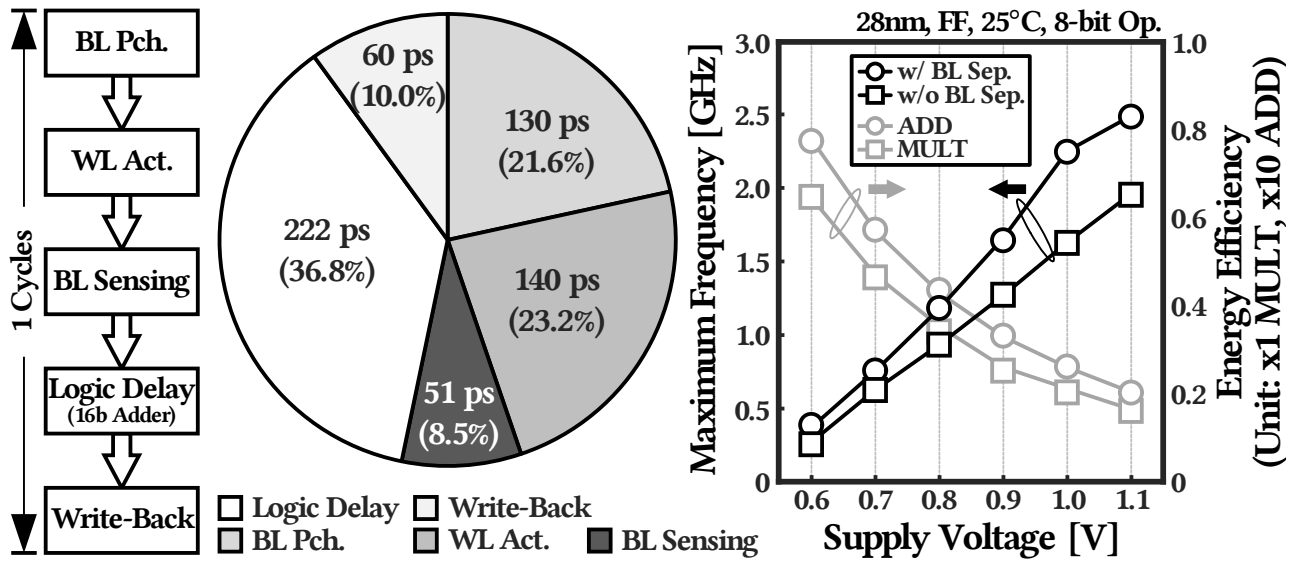

**Figure 8: Breakdown and maximum frequency of the proposed in-memory-computing.**

Fig. 6 illustrates the reconfigurable unit structure of a 2-bit precision arithmetic operation and its extension to the 4-bit and 8-bit cases. To distinguish the accessed cells from the unaccessed cells in the 4:1 column-interleaved SRAM, the accessed cells are colored grey. As shown in Fig. 6, in order to support 2-bit precision multiplication, additional 2-bit storages are needed in the proposed in-memory-computing architecture. When the bit precision of mulplication is extended two times larger, the storages also have to increase twice in their size. Thus, 2-bit FF based structure is a perfect fit for the proposed architecture, since there will be no redundant hardware in any form of operation. Our proposed architecture supports up to 8-bit precision mode, but 16-bit and 32-bit precision can also be implemented in the same method.

## 4 IMPLEMENTATION RESULT

In the proposed in-memory computing architecture, a 128KB memory size (4 banks 128 x 128 macro) has been simulated with a 28nm CMOS process. Fig. 7 (a) shows the bit-line computing delay (from WL driver to single-ended SA) compared with 0.55V WL driver based 6T SRAM in-memory computing, in various process corners. Since a short WL pulse makes small discharge and the remainder is discharged by an additional pull-down path in the BL boosting scheme, which consists of larger transistors than 6T SRAM cells, the delay of BL computing is improved 0.22X compared to the 6T SRAM with 0.55V driven WL at worst case. Also, Fig. 7 (b) shows the carry propagation delay comparison between the proposed FA and the logic gate based FA. While the proposed FA generates added results in advance and they are selected by the carry signals, logic gate based FA has to perform computation every time it receives the carry-in. Therefore, the proposed FA improves the critical path delay 1.8X-2.2X. In our proposed architecture, the overall delay consists of components as shown on the left side of Fig. 8. Due to the reduction of delay of each component, we can enhance the operating frequency. In the 8-bit precision case, we have to consider logic delay the same as that of 16-bit adder delay, which is 222ps. In addition, we can reduce the write-back delay by controlling the BL separator.

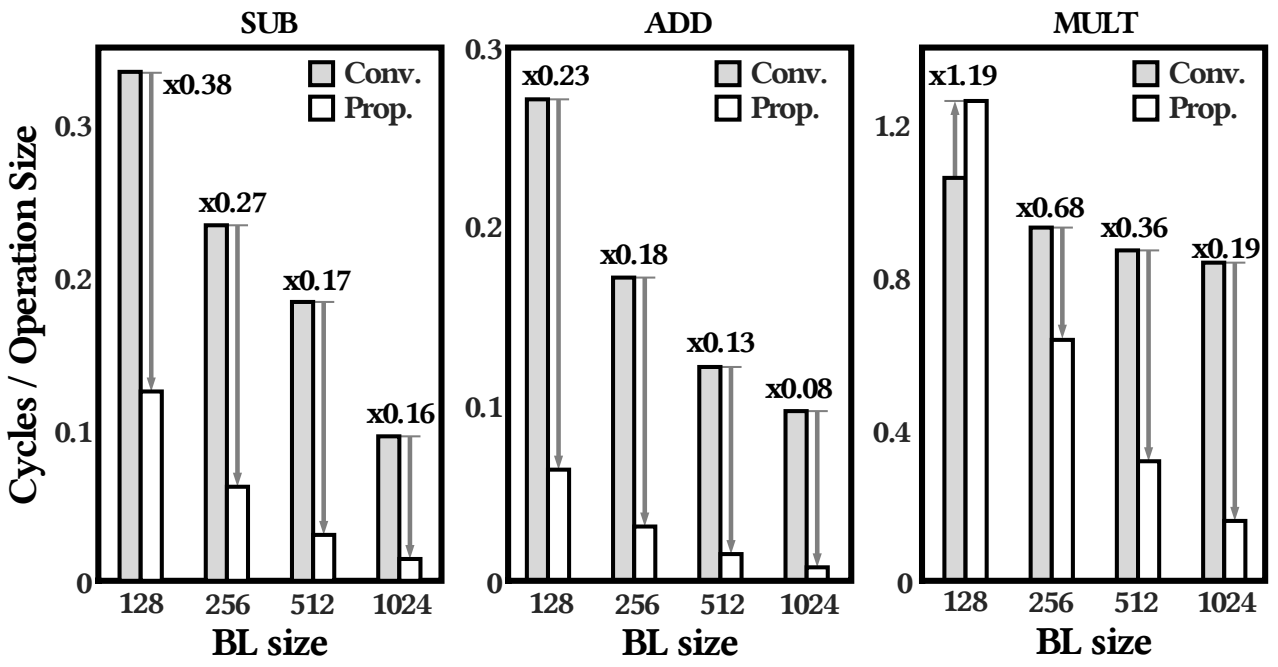

**Figure 9: Operation cycle comparison with conventional bit-serial approach [2].**

**Table II: Supported operations and their energy**

| Operation | ADD | | | SUB | | | MULT | | |
|---|---|---|---|---|---|---|---|---|---|
| Bit-Precision | 2-bit | 4-bit | 8-bit | 2-bit | 4-bit | 8-bit | 2-bit | 4-bit | 8-bit |
| Energy/ Operation [fJ] | 68.2 | 138.4 | 274.8 | w/o BL Separator | | | w/o BL Separator | | |
| | | | | 152.3 | 307.5 | 612.2 | 357.4 | 1167.6 | 4186.4 |
| | | | | w/ BL Separator | | | w/ BL Separator | | |
| | | | | 136.5 | 274.9 | 545.4 | 296 | 922.4 | 3394.8 |

The right side of Fig. 8 presents maximum operation frequency with various supply voltages, and the TOPS/W of addition/multiplication. For the figure that shows TOPS/W of adding operation, the y-axis values have to be multiplied by 10. Fig. 9 shows cycles per operation size of 8-bit arithmetic operations depending on the BL size. Compared to the conventional bit-serial approach, our bit-parallel architecture shows better cycle performance as the BL size increases because of the bit-parallel structure. Table II shows the energy per operation of typical complex computations for each bit-precision. Energy of subtraction or multiplication, which requires write-back phase, is denoted with and without the BL separator. Table III shows the comparison between the proposed and the state-of-the-arts. Our work shows the best energy efficiency (TOPS/W) with high clock frequency while avoiding read disturbance issue.

# 5 CONCLUSIONS

In this paper, we propose 6T SRAM based bit-parallel in-memory computing. By performing BL computing with a short WL pulse followed by BL boosting, we enhance the memory operation frequency while avoiding cell read disturbance issues. In addition, our transmission gate based FA composes ripple carry adder faster than logic-gate based FA. Moreover, WB delay and energy consumption are reduced due to BL separator. Due to the applied techniques, we successfully improved the memory clock frequency. Also, the bit-parallel architecture allows our IMC to perform complex operations such as addition/multiplication with

**Table III: Comparison with state-of-the-arts**

| Reference | 16' JSSC [1] | 19' JSSC [2] | 19' DAC [5] | Prop. |
|---|---|---|---|---|
| Used SRAM cell | 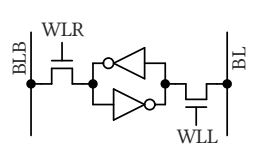 | 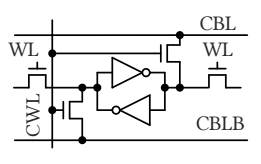 | 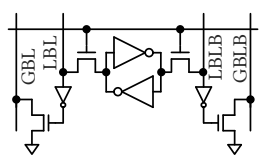 | 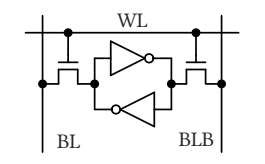 |
| Cell type | 6T cell | 8T Tranposable | 6T w/ local group | 6T cell |
| Area overhead | - | * 4.5% | * 4.0% | 5.2% |
| Read disturb | WL Underdrive | WL Underdrive | Local Read BL | Short WL w/ BL Boosting |
| Drawback | Only bit-wise operation | High latency | Local limited access | - |
| Technology | 28nm FDSOI | 28nm CMOS | 28nm CMOS | 28nm CMOS |
| Supply | 0.7V – 1.0V | 0.6V – 1.1V | 0.6V – 1.1V | 0.6V – 1.1V |
| Max Freq. | 787MHz | 475MHz (1.1V) | 2.2 GHz (1.0V) | 2.25GHz (1.0V) |
| Array size | 64 x 64 (4kB) | 4 x 128 x 256 | 256 x 128 | 4 x 128 x 128 |
| TOPS/W (MULT) | - | 0.56 (0.6V, 114MHz) | - | 0.68 (0.6V, 372MHz) |
| TOPS/W (ADD) | - | 5.27 (0.6V, 114MHz) | - | 8.09 (0.6V, 372MHz) |
| Reconfigure | X | Programmable | X | 2bit/4bit/8bit |

* Array area overhead is not included

low latency. Furthermore, this work supports reconfigurable bit-precision operation, so we can implement various algorithms and increase hardware utilization. The numerical result shows that the proposed IMC architecture achieves 2.25GHz clock frequency at 0.9V with 5.2% of area overhead. The parallel addition and multiplication of the proposed architecture also achieves 0.68, 8.09 TOPS/W for addition and multiplication, respectively. The proposed work also supports a wide range of supply voltage, from 0.6V to 1.1V.

# 6 ACKNOWLEDGEMENTS

This research was supported by the National Research Foundation of Korea grant funded by the Korea government (No. NRF-2020R1A2C3014820), and Samsung Electronics.